\documentclass[a4paper,onecolumn,11pt,unpublished]{quantumarticle}
\pdfoutput=1

\usepackage[utf8]{inputenc}
\usepackage[english]{babel}
\usepackage[T1]{fontenc}    

\usepackage{amsmath}
\usepackage{physics}        

\usepackage{hyperref}       

\usepackage{graphicx}
\usepackage{caption}
\usepackage{subcaption}     

\usepackage{quantikz}       

\usepackage[numbers]{natbib}

\begin{document}

\title{Orthogonal-Ansatz VQE: Locating excited states without modifying a cost-function}

\author[1]{Kyle Sherbert}
\orcid{0000-0002-5258-6539}
\author[1,2]{Marco Buongiorno Nardelli}
\affiliation[1]{University of North Texas, Denton, TX, 76203, USA}
\affiliation[2]{Santa Fe Institute, Santa Fe, NM, 87501, USA}
\email{mbn@unt.edu}
\maketitle

\begin{abstract}
Most literature in the Variational Quantum Eigensolver (VQE) algorithm focuses on finding the ground state of a physical system, by minimizing a quantum-computed cost-function.
When excited states are required, the cost-function is usually modified to include additional terms ensuring orthogonality with the ground state.
This generally requires additional quantum circuit executions and measurements, increasing algorithmic complexity.
Here we present a design strategy for the variational ansatz which enforces orthogonality in candidate excited states while still fully exploring the remaining subset of Hilbert space.
The result is an excited-state VQE solver which trades increasing measurement complexity for increasing circuit complexity.
The latter is anticipated to become preferable as quantum error mitigation and correction become more refined.
We demonstrate our approach with three distinct ansatze, beginning with a simple single-body example, before generalizing to accommodate the full Hilbert space spanned by all qubits, and a constrained Hilbert space obeying particle number conservation.
\end{abstract}

\begin{figure}[h]
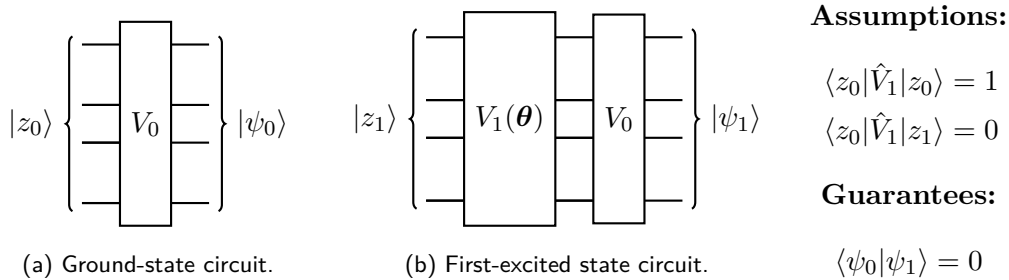

     \centering
     \begin{subfigure}[b]{0.3\textwidth}
         \centering
         \input{ct/diagram_ground}
         \caption{Ground-state circuit.}
         \label{fig:diagram:ground}
     \end{subfigure}
     \begin{subfigure}[b]{0.4\textwidth}
         \centering
         \input{ct/diagram_excited}
         \caption{First-excited state circuit.}
         \label{fig:diagram:excited}
     \end{subfigure}
     \begin{subfigure}[b]{0.2\textwidth}
         \begin{center}
             \textbf{Assumptions:}
             \begin{align*}
                \mel{z_{0}}{\hat V_1}{z_0} = 1 \\
                \mel{z_{0}}{\hat V_1}{z_1} = 0
             \end{align*}
             \textbf{Guarantees:}
             \begin{align*}
                 \braket{\psi_0}{\psi_1} = 0
             \end{align*}
         \end{center}
     \end{subfigure}
    \caption{Given a quantum circuit $V_0$ to transform a particular reference state $\ket{z_0}$ into a particular solution state $\ket{\psi_0}$ (Fig.~\ref{fig:diagram:ground}), design a quantum circuit $V_1(\vb*{\theta})$ which acts trivially on the original reference state $\ket{z_0}$ for all $\vb*{\theta}$. Applying $V_1$ and then the original circuit $V_0$ to an orthogonal reference state $\ket{z_1}$ (Fig.~\ref{fig:diagram:excited}) prepares an ansatz $\ket{\psi_1}$ {\em guaranteed} to be orthogonal to the original solution $\ket{\psi_0}$. Our paper offers several circuit families with this property and applies them to locate excited electronic states.
    }
    \label{fig:diagram}
\end{figure}

\section{Introduction}
\label{sec:intro}

Quantum chemistry and materials science are anticipated to be among the first of applications transformed by quantum computing.
Several promising algorithms have been developed over the past two decades for characterizing electronic states, including 
the Variational Quantum Eigensolver (VQE) \cite{Peruzzo_2014,McClean_2016,Cerezo_2020DEC},
Quantum Phase Estimation \cite{Abrams_1999,AsupuruGuzik_2005,Dobsicek_2007},
Quantum Subspace Expansion \cite{Huggins_2020,Klymko_2021,Manrique_2021},
Quantum Imaginary Time Evolution \cite{Motta_2020FEB,Kosugi_2021},
and many others.
See \cite{Bauer_2020} for a thorough review.

Variational algorithms like VQE are particularly popular today because their hybrid quantum-classical approach renders them compatible with the available qubit count, connectivity, decoherence, and gate fidelity, which pose severe engineering constraints on many other algorithms' efficacy.
In VQE, a parameterized quantum circuit is used to prepare an arbitrary trial state, whose energy is measured by estimating the expectation value of a molecule or material's Hamiltonian operator $\hat H$.
The parameters are then updated according to a classical optimization routine in an attempt to minimize the energy.
Despite theoretical challenges in minimizing the potential energy surface of a chemical system \cite{Cerezo_2021,Bittel_2021},
many recent experiments have reported that VQE implemented on actual quantum hardware is capable of producing a high-quality estimate of the ground state.\cite{Kandala_2017,Dumitrescu_2018,Nam_2020}

Many properties of interest require knowledge not only of the ground-state, but of excited states as well.
Most strategies for locating excited states involve a modification of the system Hamiltonian, and therefore the cost-function to be measured on the quantum computer.
For example, if one is interested in the lowest excited state above an energy $E_0$, one may minimize the energy of a ``folded'' Hamiltonian $(\hat H - E_0)^2$, which will tend to quadratically increase the complexity of the cost-function.\cite{Higgott_2019}
More commonly, one attempts to find excited states iteratively, beginning from the ground state and working up one energy level at a time.
One strategy is to directly subtract from the Hamiltonian the subspace spanned by the ground state \cite{Cerasoli_2020}, but we find this approach quickly leads to intractably complex cost-functions.
This method is refined in Orthogonally-Constrained VQE (OC-VQE): one measures the energy of a trial state as usual, but adds to the cost-function an additional term obtained from a subsequent circuit measuring orthogonality with the ground state, which ensures the trial state will find its optimal value in the excited subspace.\cite{Ryabinkin_2019,Lee_2019,Higgott_2019}
As each excited state is located, every cost-function evaluation requires running additional circuits to guarantee each trial state optimizes to the correct subspace.

In this paper, we present Orthogonal-Ansatz VQE (OA-VQE), an alternative iterative eigensolver comparable to OC-VQE.
In OA-VQE, trial-states are directly constrained to explore the excited subspace by modification of the parameterized quantum circuit rather than the cost-function.
This has the immediate benefit of keeping the cost-function consistent for each excited state optimization - measurement complexity is determined only by the complexity of the Hamiltonian itself.
Additionally, because the circuit must explore a smaller subspace each time a new excited state is located, the optimization step in OA-VQE becomes {\em easier} as the algorithm proceeds (albeit with a possibly longer circuit).

In Section~\ref{sec:design}, we describe the properties a parameterized quantum circuit family must have to be compatible with OA-VQE, and we summarize the algorithm.
In Sections~\ref{sec:reciprocal}, \ref{sec:compact}, and~\ref{sec:many}, we present circuit families suitable for performing OA-VQE for various chemical systems and qubit mappings.
In Section~\ref{sec:conclusion}, we conclude by commenting on the limitations and potential of our algorithm.

\section{Orthogonal-Ansatz VQE}
\label{sec:design}
Suppose the active space of our system of interest comprises the $N$ computational basis vectors $\ket{z_i}$, where $i\in\qty{0, 1 ... N-1}$.
Our objective is to iteratively construct an alternate basis (such as an eigenbasis for some observable) with vectors $\ket{\psi_l}\in\mathrm{Span}\qty{\ket{z_i}, i<N}$ guaranteeing orthogonality: $\braket{\psi_k}{\psi_l}=\delta_{kl}$.

Define the family of unitary operators $\hat\Omega_l$ with the following conditions:
\begin{align}
    \label{eq:omega:unitary} \hat\Omega_l \hat\Omega_l^\dagger = \hat\Omega_l^\dagger \hat\Omega_l = \hat I \\
    \label{eq:omega:inactive} \mel{z_{i<l}}{\hat\Omega_l}{z_i} = 1 \\
    \label{eq:omega:active} \mel{z_{i<l}}{\hat\Omega_l}{z_l} = 0
\end{align}
Eq.~\ref{eq:omega:unitary} is the unitary constraint.
Eq.~\ref{eq:omega:inactive} ensures that the $l$th $\hat\Omega$ acts trivially on all basis vectors indexed before $l$.
Note that Eq.~\ref{eq:omega:inactive} implies $\hat\Omega^\dagger$ must act trivially on the same vectors.
Eq.~\ref{eq:omega:active} is redundant (it can be inferred from Eqs.~\ref{eq:omega:unitary} and ~\ref{eq:omega:inactive}) but instructive: it ensures that the $l$th $\hat\Omega$ operating on the $l$th basis vector results in a vector which spans only those basis vectors indexed on or after $l$.

We now construct $\ket{\psi_l}$ as:
\begin{align}
    \label{eq:psi:l} \ket{\psi_l} = \prod_{i=0}^{l} \hat\Omega_i \ket{z_l}
\end{align}
The product is ordered such that $\hat\Omega_l$ acts directly on $\ket{z_l}$.
In quantum computing parlance, the circuit implementing $\hat\Omega_l$ is applied to the state $\ket{z_l}$, and then each preceding circuit is applied, ending with the one implementing $\hat\Omega_0$.

Consider the inner product $\braket{\psi_k}{\psi_l}$ for $k<l$:
\begin{align}
    \braket{\psi_k}{\psi_l} &= \bra{z_k} \qty( \hat\Omega_k^\dagger ... \hat\Omega_0^\dagger) \qty( \hat\Omega_0 ... \hat\Omega_k \hat\Omega_{k+1} ... \hat\Omega_l ) \ket{z_l} \nonumber \\
    &= \bra{z_k} \hat\Omega_{k+1} ... \hat\Omega_l \ket{z_l}
\end{align}
where equality is due to Eq.~\ref{eq:omega:unitary}.
Because of Eq.~\ref{eq:omega:inactive}, the bra $\bra{z_k}\hat\Omega_{k+1}$ can be contracted to $\bra{z_k}$.
Repeating this contraction, we are left with 
\begin{align}
    \braket{\psi_k}{\psi_l} = \braket{z_k}{z_l} = 0
\end{align}
Thus, the orthogonality of $\ket{\psi_l}$ is guaranteed.

Now we will describe the Orthoganal-Ansatz Variational Quantum Eigensolver (OA-VQE) algorithm.
We would like to find the eigenstates of a Hermitian operator $\hat H$ acting on a given active space.
We should construct a parameterized quantum circuit $\hat V_0(\vb*{\theta})$ such that the ansatz $\ket{\Psi(\vb*{\theta})}=\hat V_0(\vb*{\theta})\ket{z_0}$ thoroughly explores the entire active space, and variationally locate the parameters $\vb*{\theta}_0$ which minimize the expectation value $\expval{\hat H}{\Psi}$.
We identify the first of our orthogonalization operators $\Omega_0\equiv\hat V(\vb*{\theta}_0)$.
Next we should construct a new parameterized quantum circuit $\hat V_1(\vb*{\theta})$ such that $\hat V_1(\vb*{\theta})$ acts trivially on $\ket{z_0}$ (Eq.~\ref{eq:omega:inactive}), and $\ket{\Phi(\vb*{\theta})}=\hat V_1(\vb*{\theta})\ket{z_1}$ thoroughly explores the span of every basis vector in the active space {\em except} $\ket{z_0}$ (Eq.~\ref{eq:omega:active}).
A new ansatz $\Psi(\vb*{\theta})=\hat\Omega_0\ket{\Phi(\vb*{\theta})}$ is guaranteed from the arguments above to be orthogonal to the ground state; minimizing its expectation value $\expval{\hat H}{\Psi}$ therefore locates the first excited state.
If circuits can be effectively parameterized to iteratively omit each basis vector, this protocol may be repeated to obtain as many eigenstates as desired, all without needing to alter the cost-function.
As an additional benefit, one may expect each optimization to require fewer and fewer resources because subsequent eigenstates are located in a lower-dimensional space.

\section{Single-Body Circuits}
\label{sec:reciprocal}

The eigendecomposition protocol described in Section~\ref{sec:design} requires designing quantum circuits which thoroughly explore the Hilbert space spanned by certain basis vectors, while acting trivially on others.
The next three sections present examples to illustrate how such circuits may be constructed for a specific problem.
In the first example, we will briefly re-present a family of circuits used to solve the band-structure of a periodic system.

\begin{figure}[t]
     \centering
     \begin{subfigure}{0.49\textwidth}
         \centering
         \includegraphics[width=\textwidth]{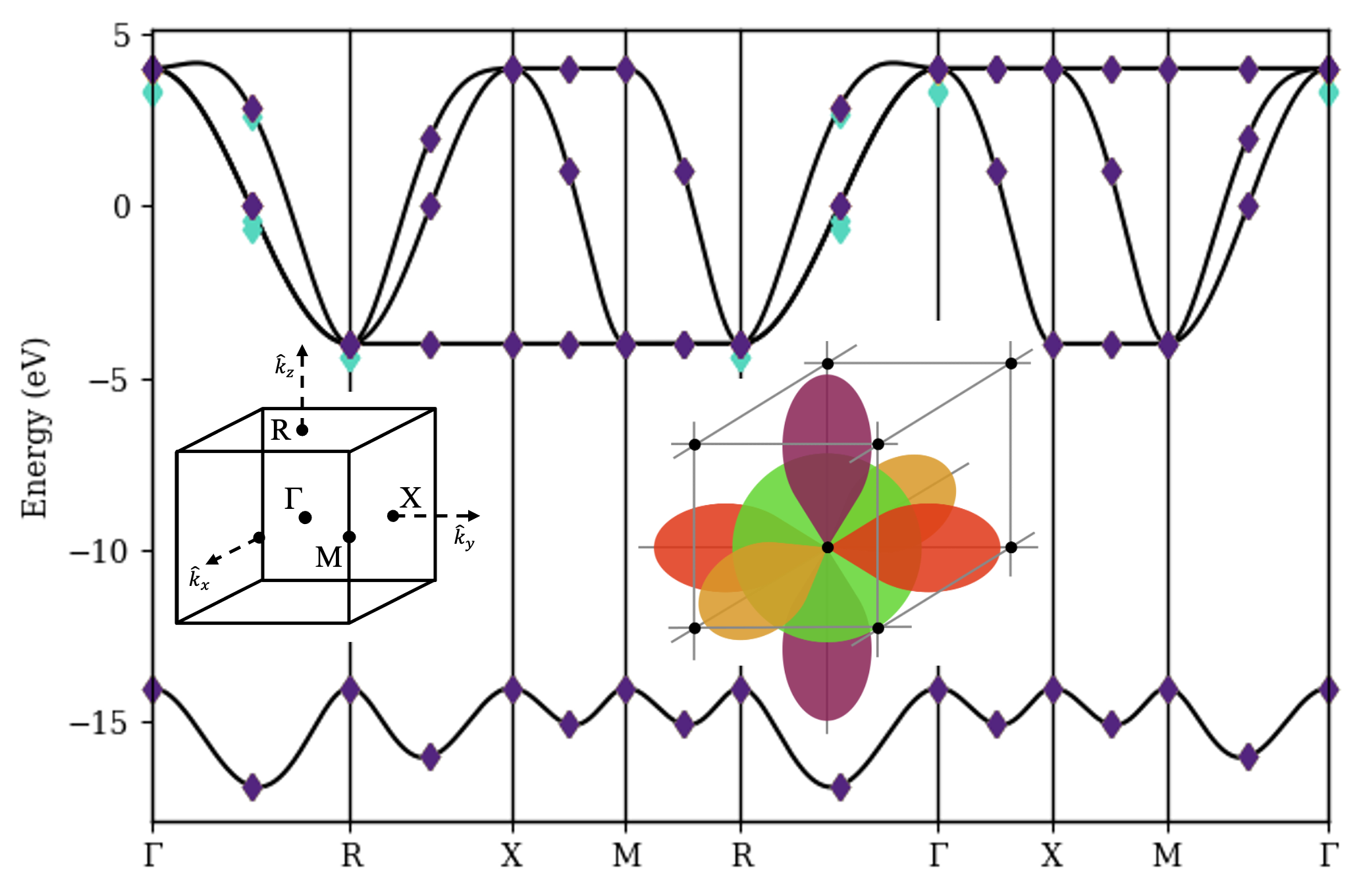}
         \caption{Toy model band structure}
         \label{fig:band:plot}
     \end{subfigure}
     \begin{subfigure}{0.49\textwidth}
         \centering
         \includegraphics[width=\textwidth]{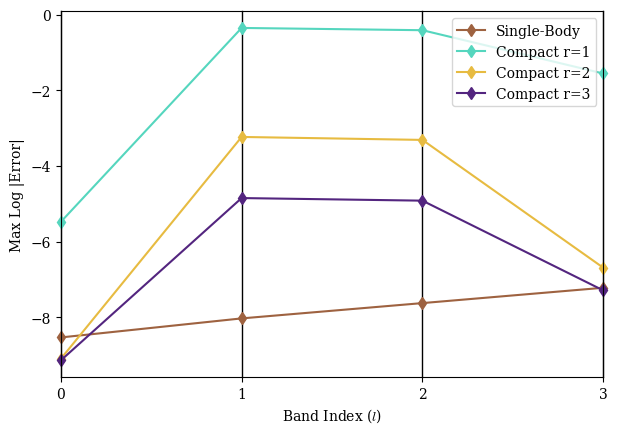}
         \caption{Worst-case optimization error.}
         \label{fig:band:error}
     \end{subfigure}
    \caption{Fig.~\ref{fig:band:plot} is the band structure and quantum variational error for a toy model of Polonium. The black curves give the analytically calculated band structure, while diamonds mark the values obtained using OA-VQE on a noiseless quantum simulator using the COBYLA optimization routine. The central inset is a schematic of the $s$ and three $p$ orbitals in a simple cubic lattice. The left inset labels the high-symmetry points in the corresponding Brillouin zone.
    Fig.~\ref{fig:band:error} marks the worst-case error for each band. The single-body circuit from Section~\ref{sec:reciprocal} is approximation-free, and optimization error is negligible. The compact circuit from Section~\ref{sec:compact} introduces approximation error via Trotterization, but this quickly falls off by Trotter-order $r=3$.}
    \label{fig:band:structure}
\end{figure}

Band-structures are a tool for characterizing periodic materials under the single-body approximation, presuming all electrons in the material behave independently.
Here we will use the reciprocal-orbital qubit mapping from \cite{Sherbert_2021}, which assigns one qubit to each orbital included in the material's unit cell.
Since we consider only one electron, the active space consists only of those basis vectors with a Hamming weight of one.
Considering a toy model for elemental polonium in a cubic lattice (Fig.~\ref{fig:band:plot}), we require four qubits; the basis vectors $\ket{1000},\ket{0100},\ket{0010},\ket{0001}$ represent the reciprocal $s$, $p_x$, $p_y$, and $p_z$ orbitals.

\begin{figure}[t]
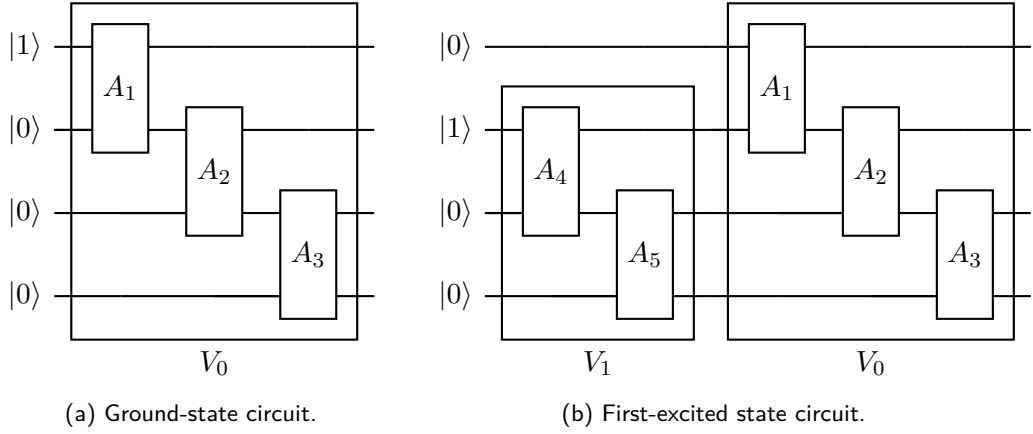

     \centering
     \begin{subfigure}[b]{0.4\textwidth}
         \centering
         \input{ct/reciprocal_ground}
         \caption{Ground-state circuit.}
         \label{fig:reciprocal:ground}
     \end{subfigure}
     \begin{subfigure}[b]{0.5\textwidth}
         \centering
         \input{ct/reciprocal_excited}
         \caption{First-excited state circuit.}
         \label{fig:reciprocal:excited}
     \end{subfigure}
    \caption{Variational circuits suitable for band-structure calculations using the reciprocal-orbital qubit mapping in Section~\ref{sec:reciprocal}. In Fig.~\ref{fig:reciprocal:ground}, all $A$ gates have two independent parameters for a total of six. In Fig.~\ref{fig:reciprocal:excited}, the $A$ gates comprising $V_0$ are fixed at their optimal values, so that only the four parameters from $V_1$ must be varied.}
    \label{fig:reciprocal}
\end{figure}

In order to construct a quantum circuit, we use the two-qubit $A$ gates presented in \cite{Gard_2020}.
Each $A$ gate takes two independent angles $\theta$, $\phi$ and rotates between the $\ket{01}$ and $\ket{10}$, such that chaining each $A$ gate together as in Fig.~\ref{fig:reciprocal:ground} successfully spans the full active space.
For excited states, cutting out individual basis vectors is as simple as cutting out qubits from the chain, as in Fig.~\ref{fig:reciprocal:excited}.
We have previously used this variational circuit design for band-structures under a different qubit mapping in \cite{Sherbert_2022}.

\section{Full Hilbert Space}
\label{sec:compact}

In this section, we consider a case when the active space corresponds to the full Hilbert space accessible by all qubits.
To ease the transition, we will continue to consider single-body band-structure calculations.
Instead of the reciprocal-orbital qubit mapping used above, we will adopt a compact mapping which assigns each orbital to its own basis state, as was done for silicon in \cite{Cerasoli_2020}.
Considering the same toy model for elemental polonium as above, we require two qubits; the basis vectors $\ket{00},\ket{01},\ket{10},\ket{11}$ represent the reciprocal $s$, $p_x$, $p_y$, and $p_z$ orbitals.

To construct a suitable variational circuit, let us begin by writing down an arbitrary wavefunction spanning the basis states $\ket{z\ge l}$:
\begin{align}
    \ket{\phi_l} = \sum_{z\ge l} \phi_z \ket{z}
\end{align}
subject to the normalization constraint $\braket{\phi_l}=1$.
Express the complex numbers $\phi_z$ in polar form as $\phi_z=r_z e^{i\gamma}$.
We facilitate parameterization of this wave-function by adopting hyperspherical coordinates $t_z, \alpha_z \in \qty[0,1]$ such that:
\begin{align}
    r_z &= \sin(\frac{\pi}{2}t_z) \prod_{i=0}^{z-1} \cos(\frac{\pi}{2}t_i) \\
    \gamma_z &= 2\pi \alpha_z
\end{align}
Normalization is enforced by setting $t_{N-1}=1$, and an arbitrary global phase is fixed by setting $\alpha_l=0$.
This leaves $2N-2$ free parameters for $\ket{\phi_0}$ to span the full Hilbert space.
Eliminating successive basis states for $\ket{\phi_{l>0}}$ is accomplished simply by setting $t_{z<l}=\alpha_{z\le l}=0$.

Now consider the following non-unitary operator:
\begin{align}
    \hat T_l = \sum_{z<l} \ketbra{z}{z} + \sum_{z\ge l} \phi_z \ketbra{z}{l}
\end{align}
It is clear that $\hat T_l \ket{l} = \ket{\phi_l}$, and that $\hat T_l$ satisfies the constraints in Eqs.~\ref{eq:omega:inactive} and~\ref{eq:omega:active}.
In order to satisfy Eq.~\ref{eq:omega:unitary} we must first construct the following Hermitian operator:
\begin{align}
    \hat H_l = \frac{1}{2} \qty(\hat T_l + \hat T_l^\dagger)
\end{align}
and unitary operator:
\begin{align}
    \hat \Omega_l = e^{i \pi \hat H_l}
\end{align}
This construction is similar in form to the popular Unitary Coupled Cluster ansatz.\cite{McClean_2016,Lee_2019}

Implementing $\hat\Omega_l$ as a quantum circuit is equivalent to time-evolving a Hamiltonian operator $\hat H$, a well-studied problem in the quantum computing literature.
The most straight-forward approach is Suzuki-Trotter decomposition \cite{Hatano_2005}, in which $\hat\Omega_l$ is approximated as the product $\hat\Omega_l \approx \qty[\prod_{k}\exp(i \frac{\pi}{r} \hat H_{lk})]^r$, where $\hat H_l=\sum_k \hat H_{lk}$ and each factor $\exp(i \frac{\pi}{r} \hat H_{lk})$ is straightforward to implement as a quantum circuit (see Appendix~\ref{sec:trotter} for a tutorial).
Error in the approximation is reduced by increasing $r$, leading to longer and longer circuits less suitable for noise-prone devices.

Fig.~\ref{fig:band:structure} demonstrates the success of band-structure calculations with self-orthogonalizing circuits on a noiseless quantum simulator.
All optimizations are carried out with the COBYLA algorithm implemented in the \verb|scipy| Python package.
Results using the single-body circuits from Section~\ref{sec:reciprocal} (four qubits) are virtually indistinguishable from the analytically obtained band-structure.
The compact basis enables the same calculation with exponentially fewer qubits (two qubits, in this case), but implementing the quantum circuit requires Trotterization, which introduces noticeable approximation error.
Nevertheless, we observe that the worst-case error falls off rapidly even for very small $r$.
That said, implementing these circuits becomes much more difficult on a real quantum device vulnerable to qubit decoherence and gate infidelity.
One possible alternative would be to exploit the Linear Combination of Unitaries lemma \cite{Berry_2015,Kalev_2021} for exponentially improved error scaling, at the cost of additional ancillary qubits.

\section{Many-body Circuits}
\label{sec:many}

In this section, we consider a standard formulation in which the active space corresponds to all many-body states with fixed particle number.
We again associate one qubit with each orbital, but we now adopt the Jordan-Wigner mapping \cite{McClean_2016,Seeley_2012} to enforce fermionic anti-commutation relations.
We shall consider the hydrogen molecule in a minimal basis (two atoms each contributing two spin-orbitals) with no symmetry reduction, so that we require four qubits.
Our active space consists of the basis vectors $\ket{1100}$, $\ket{1010}$, $\ket{1001}$, $\ket{0110}$, $\ket{0101}$, $\ket{0011}$.

\begin{figure}[t]
     \centering
     \begin{subfigure}{0.49\textwidth}
         \centering
         \includegraphics[width=\textwidth]{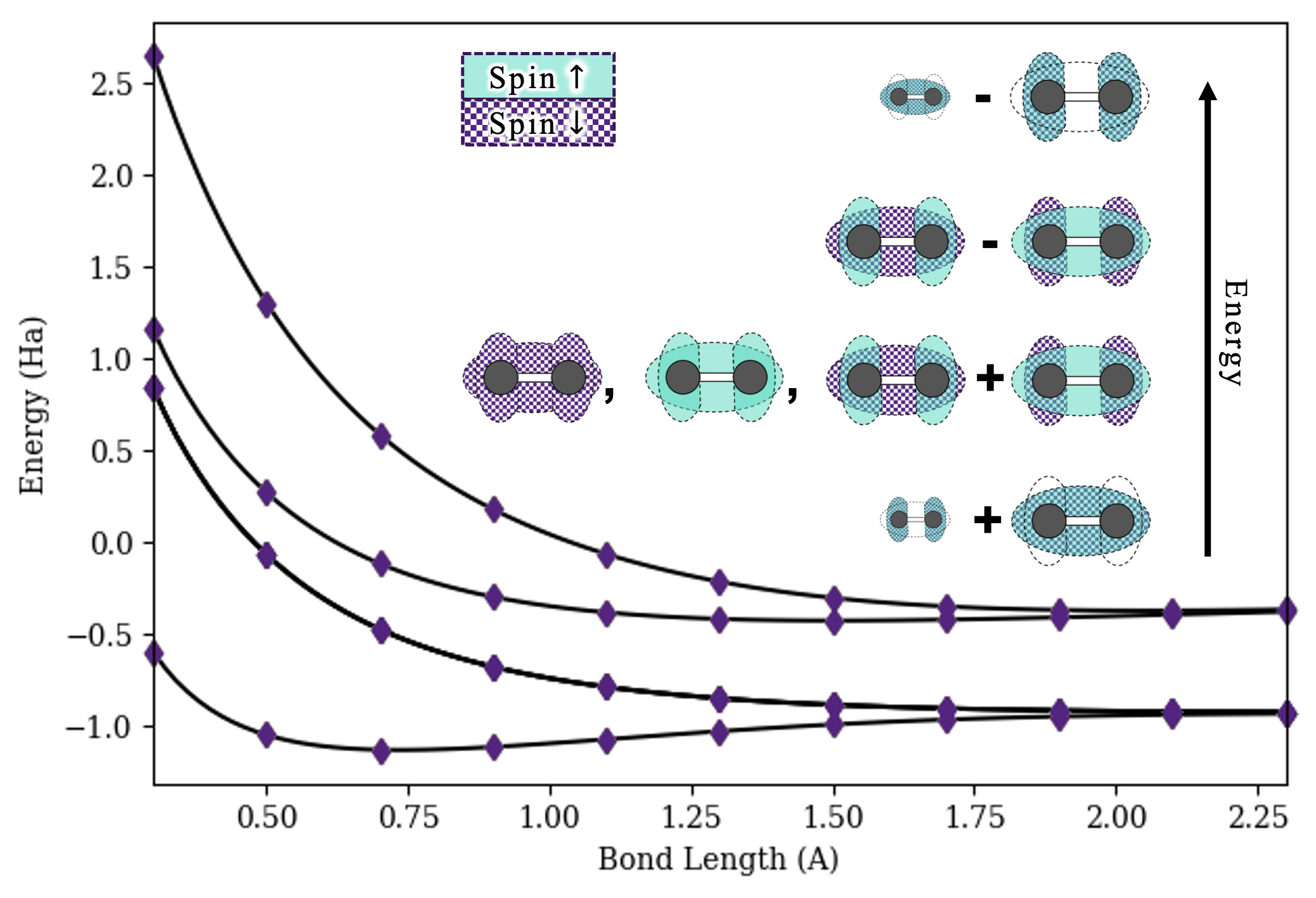}
         \caption{Hydrogen dissociation curve}
         \label{fig:hydrogen:plot}
     \end{subfigure}
     \begin{subfigure}{0.49\textwidth}
         \centering
         \includegraphics[width=\textwidth]{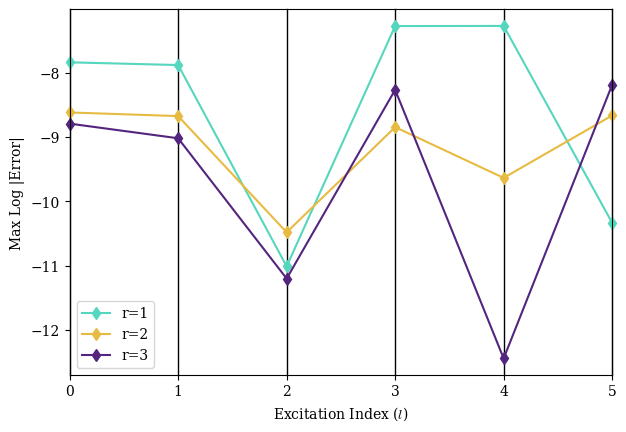}
         \caption{Worst-case optimization error.}
         \label{fig:hydrogen:error}
     \end{subfigure}
    \caption{Fig.~\ref{fig:hydrogen:plot} is the dissociation curve for the hydrogen molecule. Solid lines give the FCI energy while diamonds mark the values obtained from OA-VQE on a noiseless quantum simulator using the COBYLA optimization routine. The bottom curve gives the electronic ground-state energy under the Born-Oppenheimer approximation over increasing atomic separation. Its minimum value at approximately .74 Angstroms marks the bond of the hydrogen molecule.
    The ground-state itself is multi-reference, consisting predominantly of the Hartree-Fock bonding orbital, but increasingly large contributions from the anti-bonding orbital as the molecule is stretched.
    The second-lowest curve is triply-degenerate, including the spatially symmetric combination of bonding and anti-bonding orbitals, and both states in which the electron spins are parallel. It has no minimum value, illustrating that the triplet state of molecular hydrogen is inherently unstable.
    Fig.~\ref{fig:band:error} marks the worst-case error for each energy level. The circuit used in Section~\ref{sec:many} introduces approximation error via Trotterization, but is practically negligible even for $r=1$, on the same order as exponentiating the Hermitian operator analytically, without Trotterization.}
    \label{fig:hydrogen}
\end{figure}

The self-orthogonalizing variational circuit and parameterization we adopt is essentially the same as in Section~\ref{sec:compact}.
The crucial difference is that we replace the ket $\ket{z}$ with the state $A^\dagger_z \ket{}$.
Here $\ket{}$ is the vacuum state, $z$ indexes the vector in our active space (note that an order must be assigned to the vectors), and $A^\dagger_z$ is shorthand for the normal-ordered product of creation operators which fill up the orbitals corresponding to the bits indicated in the basis vector.
For example, if the basis vectors are ordered as in the previous paragraph, then $A^\dagger_0=a^\dagger_3 a^\dagger_2$, $A^\dagger_1=a^\dagger_3 a^\dagger_1$, and so on.
These creation operators are then mapped onto qubit operators via the Jordan-Wigner mapping, and $\hat\Omega_l$ is implemented via Suzuki-Trotter decomposition or a comparable method as described above.

Fig.~\ref{fig:hydrogen} gives the ground-state dissociation curve of the hydrogen molecule, in addition to the excited state energies for this basis.
Orbital integral calculations are made using the \verb|pyscf| Python package \cite{Sun_2020} and the Jordan-Wigner mapping is carried out with the \verb|openfermion| Python package.\cite{McClean_2020}
While we should not expect the minimal basis to yield extremely accurate calculations for excited states, we can observe the long-known fact that the first-excited state (triply-degenerate) lacks a minimum, illustrating the fact that electronic excitations in hydrogen gas tend to split the molecule apart.\cite{Berry_1980}
As observed in the compact single-body example in Fig.~\ref{fig:band:structure}, the circuits $\hat\Omega_l$ require only very small Trotterization constants $r$ to be effective.

\section{Conclusion}
\label{sec:conclusion}

In this paper, we have formulated an Orthogonal-Ansatz Variational Quantum Eigensolver (OA-VQE) algorithm for iteratively locating excited states, using variational quantum circuits spanning only the subspace orthogonal to previously-located states.
The main distinction of this approach as compared to Orthogonally-Constrained VQE (OC-VQE) is to leave the cost-function unchanged for all excited states, simplifying measurement complexity.

We have offered several families of self-orthogonalizing variational circuits, suitable for studying both single-body band structure calculations and many-body molecular spectroscopy.
However, we emphasize that the circuit families presented in Sections~\ref{sec:reciprocal}-\ref{sec:many} are by no means the only circuits consistent with the self-orthogonality constraints presented in Section~\ref{sec:design}.
If other variational circuit families (eg. UCC, ADAPT, hardware-efficient ansatze, etc.) can be adapted to be consistent with Eqs.~\ref{eq:omega:unitary}-\ref{eq:omega:active}, they will be compatible with OA-VQE.

For example, the circuit families presented in Sections~\ref{sec:compact} and~\ref{sec:many} have a similar form to Unitary Coupled Cluster (UCC) circuits.
The ``cluster operators'' $\hat T_l$ used in this paper are designed to robustly explore the entire Hilbert subspace orthogonal to previously located eigenstates.
Meanwhile, the standard UCC cluster operator is formulated to preferentially explore the basis vectors most likely contributing to the ground state, for example by considering only single- and double-excitations (UCCSD) from the Hartree-Fock solution.
Reconciling the UCCSD approximation with the self-orthogonality constraints presented here could lead to fewer terms in the Hermitian operator $\hat H_l$ and thus more efficient implementations of $\hat\Omega_l$ as a quantum circuit.

The improved measurement complexity in OA-VQE is balanced by an increased circuit complexity required to guarantee orthogonality.
In the short-term, this is a high price to pay due to relatively low coherence time and gate fidelity in present-day quantum computers.
However, the relatively high cost of measuring and resetting qubits suggests that hybrid quantum-classical algorithms such as VQE may soon transition to favoring fewer, longer circuits.
Moreover, not every circuit family will necessarily incur an asymptotic cost in circuit complexity: for example, the single-body circuits presented in Fig.~\ref{fig:reciprocal} can overlap such that the circuit length is consistently $\mathcal{O}(N)$.
This emphasizes the importance of problem-specific quantum circuit design, rather than a one-size-fits-all approach, when developing algorithms for quantum chemistry and materials science.

\section{Acknowledgments}
\label{sec:acknowledgments}
We thank Marco Fornari, Itay Hen, and Rosa Di Felice for useful discussions. We acknowledge support from the US Department of Energy through the grant \textit{Q4Q: Quantum Computation for Quantum Prediction of Materials and Molecular Properties} (DE-SC0019432).

\appendix
\section{Suzuki-Trotter Circuits}
\label{sec:trotter}
This appendix is meant for readers unfamiliar with Hamiltonian simulation via Suzuki-Trotter simulation.
We remind the reader that other methods of Hamiltonian simulation exist, such as in \cite{Berry_2015,Kalev_2021}, but we provide this tutorial so that the industrious reader may reproduce our procedure.

Our objective is to implement the unitary operator $\hat\Omega=\exp(i\pi\hat H)$ as a quantum circuit.
We first map the Hamiltonian $\hat H$ onto a sum of qubit operators:
\begin{align}
    \hat H \rightarrow \sum_{k=1}^{4^n} c_k \hat P_k
\end{align}
Each $\hat P_k$ is a ``Pauli word'', consisting of one of the Pauli-spin operators $\hat X$, $\hat Y$, $\hat Z$, or the identity operator $\hat I$, acting on each of $n$ qubits.
The coefficients $c_k$ are determined by the particular qubit mapping.
In Section~\ref{sec:compact}, we map projection operators $\ketbra{z}{z'}$ directly onto Pauli words; we refer the reader to \cite{Sherbert_2022} for details.
In Section~\ref{sec:many}, we map a second-quantized Hamiltonian onto Pauli words using the Jordan-Wigner mapping; see \cite{Seeley_2012} or documentation for the \verb|openfermion| Python package \cite{McClean_2020} for details.

Next we take advantage of the second-order exponential product formula (also known as the first-order symmetrized product formula):
\begin{align}
    e^{x(A+B+...Z)} = e^{xA/2} e^{xB/2} ... e^{xZ} ... e^{xB/2} e^{xA/2} + \mathcal{O}(x^3)
\end{align}
Higher-order formulae exist, but become needlessly complicated - see \cite{Hatano_2005} for an excellent tutorial.
Therefore, the following decomposition is sufficient:
\begin{align}
    \hat\Omega=e^{i\pi\hat H} \approx \prod_{k=1}^{4^n} e^{i\pi c_k \hat P_k/2} \prod_{k=4^n}^{1} e^{i\pi c_k \hat P_k/2}
\end{align}

If the coefficient $x$ in the exponential product formula is sufficiently small, the single-operator exponentials on the right-hand side are an adequate approximation of the complex operator exponential on the left.
In order to {\em guarantee} the operator $x$ is sufficiently small, we introduce Trotterization:
\begin{align}
    \hat\Omega \approx \qty[\prod_{k=1}^{4^n} e^{i \frac{\pi c_k}{2r} \hat P_k} \prod_{k=4^n}^{1} e^{i\frac{\pi c_k}{2r} \hat P_k}]^r
\end{align}
Let $\Lambda=\max_k c_k$; then the error in the product formula scales with $\mathcal{O}(\frac{\Lambda^3}{r^3})$.

\begin{figure}[t]
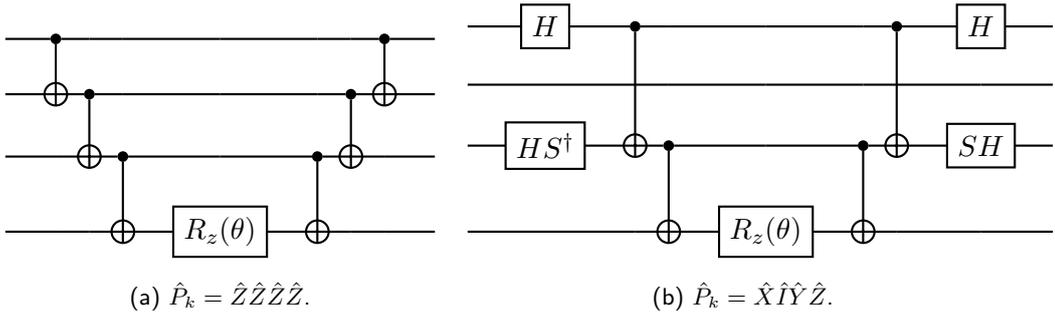

     \centering
     \begin{subfigure}[b]{0.4\textwidth}
         \centering
         \input{ct/exponential_ZZZZ}
         \caption{$\hat P_k = \hat Z\hat Z\hat Z\hat Z$.}
         \label{fig:exponential:ZZZZ}
     \end{subfigure}
     \begin{subfigure}[b]{0.5\textwidth}
         \centering
         \input{ct/exponential_XIYZ}
         \caption{$\hat P_k = \hat X\hat I\hat Y\hat Z$.}
         \label{fig:exponential:XIYZ}
     \end{subfigure}
    \caption{Circuits to implement $\hat\Omega_k = \exp(i\theta\hat P_k/2)$. $R_z$ is the standard single-qubit rotation about the $z$-axis, $H$ is the standard Hadamard gate, and $S$ is the standard phase gate. The solitary $H$ gate serves to rotate the qubits corresponding to an $\hat X$ operator to and from the $X$ basis, and the $HS^\dagger$ gate serves the same purpose for the $Y$ basis. Note that qubits corresponding to an $\hat I$ operator are ignored.}
    \label{fig:exponential}
\end{figure}

It now suffices to specify how to implement the factor $\hat\Omega_k=\exp(i\theta\hat P_k/2)$ as a quantum circuit.
The circuit for $\hat P_k=\hat Z\otimes\hat Z\otimes\hat Z\otimes\hat Z$ is given in Fig.~\ref{fig:exponential:ZZZZ}.
To account for $\hat I$ operators, one simply omits the CNOT gate on the corresponding qubit.
To account for $\hat X$ and $\hat Y$ operators, one must apply basis-rotation gates before and after: Fig.~\ref{fig:exponential:XIYZ} illustrates the circuit for $\hat P_k=\hat X\otimes\hat I\otimes\hat Y\otimes\hat Z$.
See \cite{Seeley_2012} for a more thorough explanation.
The full operator $\hat\Omega$ is constructed by applying each $\hat\Omega_k$ one after the other.

\bibliographystyle{quantum}
\bibliography{ms}

\begin{thebibliography}{10}

\bibitem{Peruzzo_2014}
Alberto Peruzzo, Jarrod McClean, Peter Shadbolt, Man~Hong Yung, Xiao~Qi Zhou,
  Peter~J. Love, Al{\'{a}}n Aspuru-Guzik, and Jeremy~L. O'Brien.
\newblock ``{A variational eigenvalue solver on a photonic quantum
  processor}''.
\newblock \href{https://dx.doi.org/10.1038/ncomms5213}{Nature Communications
  {\bf 5}, 4213}~(2014).

\bibitem{McClean_2016}
Jarrod~R. McClean, Jonathan Romero, Ryan Babbush, and Al{\'{a}}n Aspuru-Guzik.
\newblock ``{The theory of variational hybrid quantum-classical algorithms}''.
\newblock \href{https://dx.doi.org/10.1088/1367-2630/18/2/023023}{New Journal
  of Physics {\bf 18}, 23023}~(2016).
\newblock  \href{http://arxiv.org/abs/1509.04279}{arXiv:1509.04279}.

\bibitem{Cerezo_2020DEC}
M.~Cerezo, Andrew Arrasmith, Ryan Babbush, Simon~C. Benjamin, Suguru Endo,
  Keisuke Fujii, Jarrod~R. McClean, Kosuke Mitarai, Xiao Yuan, Lukasz Cincio,
  and Patrick~J. Coles.
\newblock ``{Variational quantum algorithms}''.
\newblock \href{https://dx.doi.org/10.1038/s42254-021-00348-9}{Nature Reviews
  Physics {\bf 3}, 625--644}~(2021).
\newblock  \href{http://arxiv.org/abs/2012.09265}{arXiv:2012.09265}.

\bibitem{Abrams_1999}
Daniel~S. Abrams and Seth Lloyd.
\newblock ``{Quantum algorithm providing exponential speed increase for finding
  eigenvalues and eigenvectors}''.
\newblock \href{https://dx.doi.org/10.1103/PhysRevLett.83.5162}{Physical Review
  Letters {\bf 83}, 5162--5165}~(1999).
\newblock  \href{http://arxiv.org/abs/9807070}{arXiv:9807070}.

\bibitem{AsupuruGuzik_2005}
Al{\'{a}}n Aspuru-Guzik, Anthony~D. Dutoi, Peter~J. Love, and Martin
  Head-Gordon.
\newblock ``{Simulated quantum computation of molecular energies}''.
\newblock \href{https://dx.doi.org/10.1126/science.1113479}{Science {\bf 309},
  1704--1707}~(2005).

\bibitem{Dobsicek_2007}
Miroslav Dob{\v{s}}{\'{i}}{\v{c}}ek, G{\"{o}}ran Johansson, Vitaly Shumeiko,
  and G{\"{o}}ran Wendin.
\newblock ``{Arbitrary accuracy iterative quantum phase estimation algorithm
  using a single ancillary qubit: A two-qubit benchmark}''.
\newblock \href{https://dx.doi.org/10.1103/PhysRevA.76.030306}{Physical Review
  A - Atomic, Molecular, and Optical Physics{\bf 76}}~(2007).

\bibitem{Huggins_2020}
William~J. Huggins, Joonho Lee, Unpil Baek, Bryan O'Gorman, and K.~{Birgitta
  Whaley}.
\newblock ``{A non-orthogonal variational quantum eigensolver}''.
\newblock \href{https://dx.doi.org/10.1088/1367-2630/ab867b}{New Journal of
  Physics {\bf 22}, 073009}~(2020).

\bibitem{Klymko_2021}
Katherine Klymko, Carlos Mejuto-zaera, Stephen~J Cotton, Filip Wudarski,
  Diptarka Hait, Jonathan Moussa, Wibe A~De Jong, and Norm~M Tubman.
\newblock ``{Real time evolution for ultracompact Hamiltonian eigenstates on
  quantum hardware}''~(2021).
\newblock  \href{http://arxiv.org/abs/2103.0856}{arXiv:2103.08563v1}.

\bibitem{Manrique_2021}
David~Zsolt Manrique, Irfan~T. Khan, Kentaro Yamamoto, Vijja Wichitwechkarn,
  and David~Mu{\~{n}}oz Ramo.
\newblock ``{Momentum-Space Unitary Coupled Cluster and Translational Quantum
  Subspace Expansion for Periodic Systems on Quantum Computers}''~(2020).
\newblock  \href{http://arxiv.org/abs/2008.08694}{arXiv:2008.08694}.

\bibitem{Motta_2020FEB}
Mario Motta, Chong Sun, Adrian~T.K. Tan, Matthew~J. O'Rourke, Erika Ye,
  Austin~J. Minnich, Fernando~G.S.L. Brand{\~{a}}o, and Garnet Kin~Lic Chan.
\newblock ``{Determining eigenstates and thermal states on a quantum computer
  using quantum imaginary time evolution}''.
\newblock \href{https://dx.doi.org/10.1038/s41567-019-0704-4}{Nature Physics
  {\bf 16}, 205--210}~(2020).
\newblock  \href{http://arxiv.org/abs/1901.07653}{arXiv:1901.07653}.

\bibitem{Kosugi_2021}
Taichi Kosugi, Yusuke Nishiya, and Yu-ichiro Matsushita.
\newblock ``{Probabilistic imaginary-time evolution by using forward and
  backward real-time evolution with a single ancilla: first-quantized
  eigensolver of quantum chemistry for ground states}''~(2021).
\newblock  \href{http://arxiv.org/abs/2111.12471}{arXiv:2111.12471}.

\bibitem{Bauer_2020}
Bela Bauer, Sergey Bravyi, Mario Motta, and Garnet {Kin-Lic Chan}.
\newblock ``{Quantum Algorithms for Quantum Chemistry and Quantum Materials
  Science}''.
\newblock \href{https://dx.doi.org/10.1021/acs.chemrev.9b00829}{Chemical
  Reviews {\bf 120}, 12685--12717}~(2020).
\newblock  \href{http://arxiv.org/abs/2001.03685}{arXiv:2001.03685}.

\bibitem{Cerezo_2021}
M.~Cerezo, Akira Sone, Tyler Volkoff, Lukasz Cincio, and Patrick~J. Coles.
\newblock ``{Cost function dependent barren plateaus in shallow parametrized
  quantum circuits}''.
\newblock \href{https://dx.doi.org/10.1038/s41467-021-21728-w}{Nature
  Communications{\bf 12}}~(2021).
\newblock  \href{http://arxiv.org/abs/2001.00550}{arXiv:2001.00550}.

\bibitem{Bittel_2021}
Lennart Bittel and Martin Kliesch.
\newblock ``{Training Variational Quantum Algorithms Is NP-Hard}''.
\newblock \href{https://dx.doi.org/10.1103/physrevlett.127.120502}{Physical
  Review Letters {\bf 127}, 120502}~(2021).

\bibitem{Kandala_2017}
Abhinav Kandala, Antonio Mezzacapo, Kristan Temme, Maika Takita, Markus Brink,
  Jerry~M. Chow, and Jay~M. Gambetta.
\newblock ``{Hardware-efficient variational quantum eigensolver for small
  molecules and quantum magnets}''.
\newblock \href{https://dx.doi.org/10.1038/nature23879}{Nature {\bf 549},
  242--246}~(2017).
\newblock  \href{http://arxiv.org/abs/1704.05018}{arXiv:1704.05018}.

\bibitem{Dumitrescu_2018}
E.~F. Dumitrescu, A.~J. McCaskey, G.~Hagen, G.~R. Jansen, T.~D. Morris,
  T.~Papenbrock, R.~C. Pooser, D.~J. Dean, and P.~Lougovski.
\newblock ``{Cloud Quantum Computing of an Atomic Nucleus}''.
\newblock \href{https://dx.doi.org/10.1103/PhysRevLett.120.210501}{Physical
  Review Letters {\bf 120}, 210501}~(2018).
\newblock  \href{http://arxiv.org/abs/1801.03897}{arXiv:1801.03897}.

\bibitem{Nam_2020}
Yunseong Nam, Jwo~Sy Chen, Neal~C. Pisenti, Kenneth Wright, Conor Delaney,
  Dmitri Maslov, Kenneth~R. Brown, Stewart Allen, Jason~M. Amini, Joel
  Apisdorf, Kristin~M. Beck, Aleksey Blinov, Vandiver Chaplin, Mika
  Chmielewski, Coleman Collins, Shantanu Debnath, Andrew~M. Ducore, Kai~M.
  Hudek, Matthew Keesan, Sarah~M. Kreikemeier, Jonathan Mizrahi, Phil Solomon,
  Mike Williams, Jaime~David Wong-Campos, Christopher Monroe, and Jungsang Kim.
\newblock ``{Ground-state energy estimation of the water molecule on a trapped
  ion quantum computer}''.
\newblock npj Quantum Information{\bf 6}~(2019).
\newblock  \href{http://arxiv.org/abs/1902.10171}{arXiv:1902.10171}.

\bibitem{Higgott_2019}
Oscar Higgott, Daochen Wang, and Stephen Brierley.
\newblock ``{Variational quantum computation of excited states}''.
\newblock \href{https://dx.doi.org/10.22331/q-2019-07-01-156}{Quantum {\bf 3},
  156}~(2019).
\newblock  \href{http://arxiv.org/abs/1805.08138}{arXiv:1805.08138}.

\bibitem{Cerasoli_2020}
Frank~T. Cerasoli, Kyle Sherbert, Jagoda S{\l}awi{\'{n}}ska, and Marco
  {Buongiorno Nardelli}.
\newblock ``{Quantum computation of silicon electronic band structure}''.
\newblock \href{https://dx.doi.org/10.1039/d0cp04008h}{Physical Chemistry
  Chemical Physics {\bf 22}, 21816--21822}~(2020).
\newblock  \href{http://arxiv.org/abs/2006.03807}{arXiv:2006.03807}.

\bibitem{Ryabinkin_2019}
Ilya~G. Ryabinkin, Scott~N. Genin, and Artur~F. Izmaylov.
\newblock ``{Constrained Variational Quantum Eigensolver: Quantum Computer
  Search Engine in the Fock Space}''.
\newblock \href{https://dx.doi.org/10.1021/acs.jctc.8b00943}{Journal of
  Chemical Theory and Computation {\bf 15}, 249--255}~(2019).
\newblock  \href{http://arxiv.org/abs/1806.00461}{arXiv:1806.00461}.

\bibitem{Lee_2019}
Joonho Lee, William~J. Huggins, Martin Head-Gordon, and K.~Birgitta Whaley.
\newblock ``{Generalized Unitary Coupled Cluster Wave functions for Quantum
  Computation}''.
\newblock \href{https://dx.doi.org/10.1021/acs.jctc.8b01004}{Journal of
  Chemical Theory and Computation {\bf 15}, 311--324}~(2019).
\newblock  \href{http://arxiv.org/abs/1810.02327}{arXiv:1810.02327}.

\bibitem{Sherbert_2021}
Kyle Sherbert, Frank Cerasoli, and Marco~Buongiorno Nardelli.
\newblock ``{A systematic variational approach to band theory in a quantum
  computer}''.
\newblock \href{https://dx.doi.org/10.1039/d1ra07451b}{RSC Adv. {\bf 11},
  39438--39449}~(2021).
\newblock  \href{http://arxiv.org/abs/2104.03409}{arXiv:2104.03409}.

\bibitem{Gard_2020}
Bryan~T. Gard, Linghua Zhu, George~S. Barron, Nicholas~J. Mayhall, Sophia~E.
  Economou, and Edwin Barnes.
\newblock ``{Efficient symmetry-preserving state preparation circuits for the
  variational quantum eigensolver algorithm}''.
\newblock \href{https://dx.doi.org/10.1038/s41534-019-0240-1}{npj Quantum
  Information{\bf 6}}~(2020).
\newblock  \href{http://arxiv.org/abs/1904.10910}{arXiv:1904.10910}.

\bibitem{Sherbert_2022}
Kyle Sherbert, Anooja Jayaraj, and Marco~Buongiorno Nardelli.
\newblock ``{Quantum algorithm for band structures with local tight-binding
  orbitals}''.
\newblock \href{https://dx.doi.org/10.21203/rs.3.rs-1318951/v1}{In
  Review}~(2022).

\bibitem{Hatano_2005}
Naomichi Hatano and Masuo Suzuki.
\newblock ``{Finding Exponential Product Formulas of Higher Orders}''.
\newblock In Arnab Das and Bikas {K Chakrabarti}, editors, Quantum Annealing
  and Other Optimization Methods.
\newblock \href{https://dx.doi.org/10.1007/11526216_2}{Pages 37--68}.
\newblock Springer Berlin Heidelberg~(2005).
\newblock  \href{http://arxiv.org/abs/0506007}{arXiv:0506007}.

\bibitem{Berry_2015}
Dominic~W. Berry, Andrew~M. Childs, Richard Cleve, Robin Kothari, and
  Rolando~D. Somma.
\newblock ``{Simulating hamiltonian dynamics with a truncated taylor series}''.
\newblock \href{https://dx.doi.org/10.1103/PhysRevLett.114.090502}{Physical
  Review Letters{\bf 114}}~(2015).
\newblock  \href{http://arxiv.org/abs/1412.4687}{arXiv:1412.4687}.

\bibitem{Kalev_2021}
Amir Kalev and Itay Hen.
\newblock ``{Quantum algorithm for simulating hamiltonian dynamics with an
  off-diagonal series expansion}''.
\newblock \href{https://dx.doi.org/10.22331/Q-2021-04-08-426}{Quantum {\bf 5},
  1--24}~(2021).
\newblock  \href{http://arxiv.org/abs/2006.02539}{arXiv:2006.02539}.

\bibitem{Seeley_2012}
Jacob~T. Seeley, Martin~J. Richard, and Peter~J. Love.
\newblock ``{The Bravyi-Kitaev transformation for quantum computation of
  electronic structure}''.
\newblock \href{https://dx.doi.org/10.1063/1.4768229}{Journal of Chemical
  Physics {\bf 137}, 224109}~(2012).
\newblock  \href{http://arxiv.org/abs/1208.5986}{arXiv:1208.5986}.

\bibitem{Sun_2020}
Qiming Sun, Xing Zhang, Samragni Banerjee, Peng Bao, Marc Barbry, Nick~S.
  Blunt, Nikolay~A. Bogdanov, George~H. Booth, Jia Chen, Zhi~Hao Cui, Janus~J.
  Eriksen, Yang Gao, Sheng Guo, Jan Hermann, Matthew~R. Hermes, Kevin Koh,
  Peter Koval, Susi Lehtola, Zhendong Li, Junzi Liu, Narbe Mardirossian,
  James~D. McClain, Mario Motta, Bastien Mussard, Hung~Q. Pham, Artem Pulkin,
  Wirawan Purwanto, Paul~J. Robinson, Enrico Ronca, Elvira~R. Sayfutyarova,
  Maximilian Scheurer, Henry~F. Schurkus, James~E.T. Smith, Chong Sun, Shi~Ning
  Sun, Shiv Upadhyay, Lucas~K. Wagner, Xiao Wang, Alec White, James~Daniel
  Whitfield, Mark~J. Williamson, Sebastian Wouters, Jun Yang, Jason~M. Yu,
  Tianyu Zhu, Timothy~C. Berkelbach, Sandeep Sharma, Alexander~Yu Sokolov, and
  Garnet Kin~Lic Chan.
\newblock ``{Recent developments in the P y SCF program package}''.
\newblock \href{https://dx.doi.org/10.1063/5.0006074}{Journal of Chemical
  Physics{\bf 153}}~(2020).
\newblock  \href{http://arxiv.org/abs/2002.12531}{arXiv:2002.12531}.

\bibitem{McClean_2020}
Jarrod~R. McClean, Nicholas~C. Rubin, Kevin~J. Sung, Ian~D. Kivlichan, Xavier
  Bonet-Monroig, Yudong Cao, Chengyu Dai, E.~Schuyler Fried, Craig Gidney,
  Brendan Gimby, Pranav Gokhale, Thomas Haner, Tarini Hardikar, Vojt{\v{e}}ch
  Havl{\'{i}}{\v{c}}ek, Oscar Higgott, Cupjin Huang, Josh Izaac, Zhang Jiang,
  Xinle Liu, Sam Mcardle, Matthew Neeley, Thomas O'Brien, Bryan O'Gorman, Isil
  Ozfidan, Maxwell~D. Radin, Jhonathan Romero, Nicolas~P.D. Sawaya, Bruno
  Senjean, Kanav Setia, Sukin Sim, Damian~S. Steiger, Mark Steudtner, Qiming
  Sun, Wei Sun, Daochen Wang, Fang Zhang, and Ryan Babbush.
\newblock ``{OpenFermion: The electronic structure package for quantum
  computers}''.
\newblock \href{https://dx.doi.org/10.1088/2058-9565/ab8ebc}{Quantum Science
  and Technology{\bf 5}}~(2020).
\newblock  \href{http://arxiv.org/abs/1710.07629}{arXiv:1710.07629}.

\bibitem{Berry_1980}
R.~Stephen Berry, Stuart~A. Rice, and John Ross.
\newblock ``{Physical Chemistry}''.
\newblock John Wiley \& Sons, Inc. ~(1980).

\end{thebibliography}

\end{document}